\documentclass[aps,pra,twocolumn,preprintnumbers,amsmath,amssymb,footinbib]{revtex4-1}
\usepackage{listings}
\usepackage{xcolor}
\usepackage{hyperref}
\usepackage{graphicx}
\usepackage{graphicx}
\usepackage{amssymb}
\usepackage{color}

\begin{document}

\title{Tangle Centric Networking}

\author{Christopher Scherb}
\email{christopher.scherb@unibas.ch}
\affiliation{University of Basel, Basel, Switzerland}

\author{Dennis Grewe}
\email{dennis.grewe@de.bosch.com}
\affiliation{Robert Bosch GmbH, Renningen, Germany}

\author{Christian Tschudin}
\email{christian.tschudin@unibas.ch}
\affiliation{University of Basel, Basel, Switzerland}


\begin{abstract}
Today's Internet is heavily used for multimedia streaming from cloud backends, while the Internet of Things (IoT) reverses the traditional data flow, with high data volumes produced at the network edge.
Information Centric Networking (ICN) advocates against a host-centric communication model which is promising for distributed edge computing environments and the execution of IoT applications in a decentralized fashion. 
By using naming schemes, data is tightly coupled to names instead of hosts which simplifies discovery and access to data and services. 
However, the tight coupling challenges network performance due to additional synchronization overhead of large data volumes and services.

We present \emph{Tangle Centric Networking (TCN)} -- a decentralized data structure for coordinated distributed applications and data exchange following principles of ICN.
TCN can react on data and service changes and update them accordingly in network nodes, provide distributed data structures and enable cooperative work on the same data without huge overhead by using Tangles for coordination. We implemented TCN in simulations and evaluated the concept against a base line scenario.
\end{abstract}

\maketitle




\section{Introduction}
\label{sec:intro}



The Internet of Things (IoT) is reversing the data flow from traditional \emph{cloud-to-edge} to \emph{edge-to-cloud} with much of the data produced at the edge.
While the number of IoT devices will double until 2025 up to 75 billion~\cite{Statista:IoT:2025}, the data produced by these devices at the edge require processing upfront before sending it towards cloud backends (e.g.,~\cite{Moustafa:rCDN:2017}).
Such a paradigm shift enables novel classes of distributed applications which are deployed in challenging environments such as highly connected automated driving, manufacturing automation or gaming.
Examples include applications demanding for timely information sharing, often to multiple consumers, which cannot be satisfied by classic cloud-based solutions.

The Edge Computing (EC) paradigm is concerned about to supporting these classes of applications. 
EC brings computational resources, storage and services from the cloud backend to the edge of the network and therefore, closer to the data origin and its consumers (e.g.,~\cite{Bonomi:2012}).
However, most of today's network architectures are still based on a host-oriented communication model, interconnecting components with each other such as the Internet Protocol (IP) technology stack.
The model challenges the network at many levels such as management and orchestration of data and services, i.e., in mobile scenarios. 
As a result, additional frameworks are introduced to manage the overhead in the communication infrastructure such as ETSI OSM~\cite{ETSI:OSM:2021} for management of services in upcoming 5G cellular deployments.
Instead, native network-layer resolution of desired data and services avoids additional communication overhead (e.g., DNS) and has the potential to reduce response times significantly. 

Information Centric Networks (ICN) present a promising solution for novel classes of distributed applications in IoT deployments (e.g.,~\cite{Amadeo:ICN:IoT:2016}).
Based on a loosely coupled communication model, ICNs use location-independent, unique \emph{content identifiers} such as naming schemes for discovering and accessing data in the network~\cite{jacobson2009networking}.
This approach couples data to an identifier instead of the host and thus avoids resolution conflicts.
It allows mobility support by nature, while not maintaining network addresses of hosts and also facilitates additional features such as in-network processing and caching of data.
Therefore, it can be a great fit to provider-agnostic distributed applications in IoT scenarios.

As ICNs target to support a wide range of distributed applications, there are special classes of application data which challenges data discovery and management,
for example, data changing at frequent intervals (e.g., a temperature sensor) as well as large volumes of data (e.g., video stream of a surveillance camera).
In order to simplify the access, ICNs propose the concept of Manifests (e.g., FLIC~\cite{tschudin2016file}). 
Manifests describe properties of a data item including a list of \emph{content identifiers} for smaller application specific chunks.

While Manifests simplify the access to special classes of application data, it introduces a challenge on the handling frequent changing data items.
Every time a data item changes, a new version of the item is created and published under a new unique \emph{content identifier}.
However, the tight coupling of data to immutable content identifiers results in frequent re-creation of the entire Manifest.
This design decision introduces the problem on how to select and announce new versions of data in the network to avoid collisions.
It either requires centralized name resolution systems managing these identifiers (e.g., NDNS~\cite{afanasyev2017ndns}), mechanisms to achieve consensus about identifiers (e.g., synchronization across nodes), or the same data is published using different identifiers, unnecessarily occupying resources, and thus, decreases network efficiency.

To address these limitations, distributed data structures offer data replication across the network. 
Examples include Distributed Hash Tables or Conflict-free replicated data types (CRDTs)~\cite{Shapiro:CRDT:2011}.
In IoT scenarios, distributed ledger technologies have gained attention in the research community in recent years.
An example to highlight is IOTA~\cite{IOTA:2021}, offering a lightweight solution to represent data in form of Tangles. 

In this paper, we propose Tangle Centric Networking (TCN) -- a network level consensus and synchronisation extensions for ICNs using Tangles~\cite{popov2018tangle}. 
Implement as directed acyclic graph (DAG), Tangles are aligned to the proposed structure of existing Manifest solutions like FLIC~\cite{tschudin2016file}, while promising additional features such as extensibility and distributed consensus.

The contributions of the paper are:
\begin{itemize}
    \item introduction of a concept of Tangles in ICNs for network level consensus and efficient management of data objects
    \item implementation and evaluation of TCN using Named Data Networking (NDN)~\cite{zhang2014named} and the PSync~\cite{Zhang:PSync:2017} protocol showing performance improvement in data retrieval
    \item discussion of further potentials and open research directions
\end{itemize}

The remainder of the paper is structured as follows: Section~\ref{sec:use_cases} introduces a use case scenario. 
Section~\ref{sec:background} provides the related work, while Section~\ref{sec:tangles} introduces the concept of TCN and certain architectural aspects in Section~\ref{sec:tangle_sync}.
Section~\ref{sec:evaluation} presents the implementation and evaluation of TCN, followed by a discussion on the results and open research directions in Section~\ref{sec:discussion}.
Finally, we conclude the paper in Section~\ref{sec:conclusion}.

\section{Use Case Scenarios}
\label{sec:use_cases}

Distributed edge computing environments enable novel classes of application scenarios in different IoT domains. 
Examples include smart factories, healthcare systems, or connected and automated driving (e.g.,~\cite{Buyya:IoT:use_cases}).
In this paper, we consider a use case from the domain of connected and automated driving, based on the special characteristics of the scenario. 

\begin{figure}
    \centering
    \includegraphics[width=.5\textwidth]{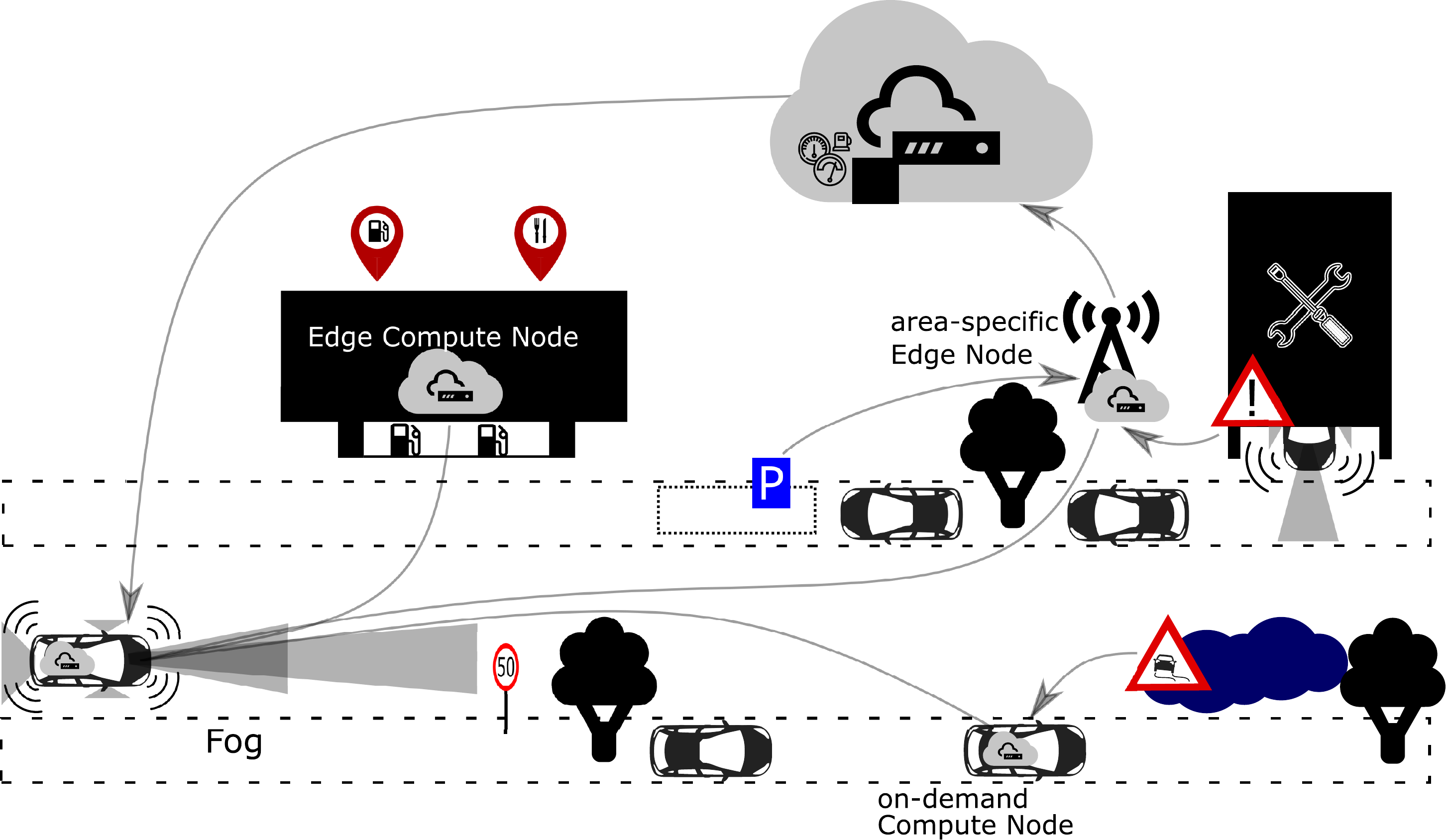}
    \caption{The \emph{electronic horizon} in a distributed edge environment. Data from different sources are collected and processed close to the origin, creating an environmental model used by automated driving logic.}
    \label{fig:electronic_horizon}
\end{figure}

\subsection{Automotive: Electronic Horizon for Highly Automated Driving}
\label{sec:use_cases:automotive}

It is evident that automated driving will rely on information from in-vehicle systems as well as data sources deployed outside the vehicle~\cite{Wang:2019}.
It is expected that driving logic in automated driving vehicles will take in several inputs ranging from built-in sensor data streams to context and external information (incl. road topology, localization, position of obstacles and moving objects, odometry, acceleration data, traffic volumes, etc.). 
The characteristics of such data are quite different in terms of volumes, popularity as well as validity, most of it generated at the edge of the communication infrastructure (cf.~\cite{Moustafa:rCDN:2017}).

Based on all information, an environmental model is created periodically -- e.g., the Local Dynamic Map (LDM)~\cite{Andreone:LDM:2010} -- serving as an input to the driving logic to further improve the driving quality.
Finally, the driving logic computes the output for steering the vehicle (steering angle, inputs to the gas \& brake pedals, switching signal lamps etc.)

On the road towards fully automated driving, the \emph{electronic horizon} describes a purely cloud-based virtual vehicle sensor using pre-defined data sources. 
These sources include map data, vehicle’s mobility model as well as additional data regarding the stretch of road ahead during a journey. 
As it is already available as a product for certain high class vehicle models (e.g., \cite{dSpace:EH:2010, Bosch:cHorizon:2020}), it forms a base line scenario for upcoming driving logic in automated driving vehicles.

Figure~\ref{fig:electronic_horizon} illustrates the electronic horizon in a distributed edge deployment. 
The fusion of all data and computation of the model require a high amount of computing power. 
Instead of sending all the data towards cloud backends, data can be collected, processed and augmented with other information by compute nodes on the delivery path close to its origin (cf.~\cite{grewe2017information}). 
Such approach is promising to reduce the massive amount of data to be transmitted towards cloud backends (see reverse CDN~\cite{Moustafa:rCDN:2017}), while increasing the overall quality of information for such scenario (e.g., enhance information from camera streams nearby).

\subsection{Challenges from the Use Case}

The presented use case introduces several challenges. 
The data to be collected is of different types (e.g., large vs. small volumes, popularity, etc.), while dynamically generated content from compute nodes challenges efficient dissemination.
From a communication and computing infrastructure perspective, data flows are reversing from downstream of cloud backends, towards upstream from devices close to the vehicle. The scenario challenges to \emph{understand} data points as well as to put data and computations into the right context (linking together) to be processed at the network edge.
From a vehicle perspective, the scenario challenges dynamically generated data and discovers the \emph{closest} compute node by taking the mobility aspect of the vehicle into account.


\section{Background \& Related Work}
\label{sec:background}

As we move toward a distributed edge computing environment, computation-centric network architecture defines a need to support decentralized and distributed computations.
The following sections will present some background and related work. 

\subsection{Information Centric Networking}
\label{sec:background:icn}
ICN describes a paradigm which puts data as the first class citizen in the network by separating content from its physical location.
We refer the term ICN to the network principles introduced by Jacobson et al.~\cite{jacobson2009networking} including architectures such as Named Data Networking (NDN)~\cite{zhang2014named}. 
By changing the addressing scheme using content identifiers (e.g., hierarchical naming schemes), ICN achieves a loosely-coupled communication model directly on the network layer.
This is in stark contrast to host-centric deployments where devices need to connect to specific nodes using their IP address. 
Therefore, ICNs are promising for data transport in deployments at the network edge as well as in IoT environments as there is no need for the network to track the location of individual nodes~\cite{Amadeo:ICN:IoT:2016, grewe2017information}.


Content in ICN is represented by so-called \emph{Named Data Objects} (NDO) that have identifiers, but that are usually not associated to a single or specific physical node in the network.
To transfer large volumes of data in an ICN, NDOs are structured into smaller chunks to fit into the maximum transmission unit of the underlying network. 
ICNs provide several solutions to discover the identifiers of chunks of a particular NDO including Manifest (e.g., FLIC~\cite{tschudin2016file}) or DNS-like resolution system such as NDNS~\cite{afanasyev2017ndns}. 
For example, FLIC is build like a UNIX directory. From a root manifest (in UNIX terms: root-directory) it points to chunks (in unix terms: files) and to the next manifests (in UNIX terms: directories). Each manifest can contain further manifests or chunks. Thus, FLIC represents a tree, where manifests are inner nodes and chunks are trees. 
This is a very efficient data structure to find certain chunks, since the search time is reduces from linear to logarithmic.
However, to support the execution of distributed application on top of ICNs, any changes to data requires updates of the NDO descriptions to provide access to vital data. 
For example, FLIC Manifests often have to be re-created entirely every time a change to a NDO happens, since each of the manifests contains signatures of the chunks/other manifests it is pointing to.

Addressing this issue, distributed data synchronization protocols have been proposed in the literature to support multi-party communication in ICNs (cf.~\cite{Shang:NDNSync:2017}).
Examples include ChronoSync~\cite{Zhu:ChronoSync:2013}, VectorSync or PSync~\cite{Zhang:PSync:2017} featuring different design rationals in naming schemes, and state propagation. However, as we move towards edge computing environment, these solutions have not been evaluated in the context of compute-centric networking architectures.

\subsection{Distributed Storage \& Ledger}
\label{sec:background:dlt}

In distributed systems, the CAP theorem (also called Brewer’s Theorem) poses that a system can have only two of three desirable properties: \emph{consistency}, \emph{availability}, and \emph{partition tolerance}~\cite{Gilbert:CAP:2002}.
However, trade-offs can be achieved by explicitly handling partition tolerance in order to optimize consistency.
For example, CFN uses CRDTs to describe the compute graph the platform should execute~\cite{krol2019compute}.
However, CRDTs only guarantee \emph{eventual consistency} demanding for global consistency - the entire data structure has to be updated in case of changes~\cite{Shapiro:CRDT:2011}.
In recent years, distributed ledger describes a technological infrastructure representing consensus by allowing to access, validate and record data across multiple parties in a distributed fashion~\cite{swan2015blockchain}.
The Blockchain~\cite{swan2015blockchain} implements a distributed, append-only structure consisting of connected blocks. 
Examples based on the technology include Bitcoin~\cite{nakamoto2019bitcoin} or Ethereum~\cite{Ethereum}.
In the context of compute-centric architectures, the SPOC~\cite{krol2018spoc} approach uses Ethereum to record outsource computations, executed by 3rd party nodes in the infrastructure, able to validate changes.

Instead of recording entries as an append-only structure of connected blocks, the Tangle~\cite{popov2018tangle} technology, which is the data structure behind IOTA~\cite{IOTA:2021}, implements a directed acyclic graph (DAG) for storing transactions.
Every new transaction has to approve at least two previous transactions to be recorded in the Tangle. A tip is a not yet approved block. Thereby, the \emph{tips selection algorithm} chooses two tips that should been approved. The 
algorithm makes sure that tips are chosen in a way, that the tangle will reach convergence quickly.

\begin{figure}
    \centering
    \includegraphics[width=.5\textwidth]{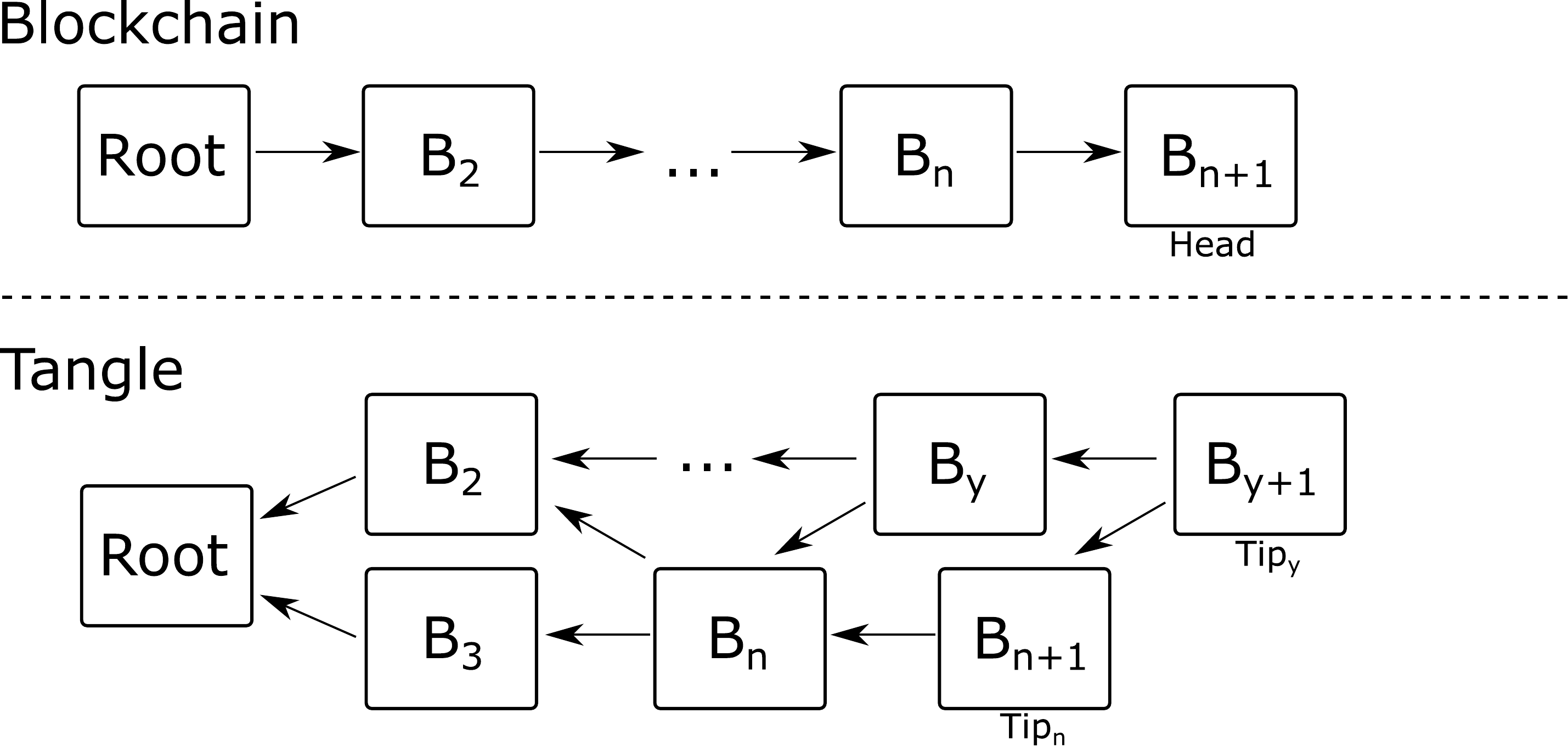}
    \caption{Illustration of the characteristics of the distributed, append-only data structure of Blockchain and Tangles.}
    \label{fig:blockchain:tangle}

\end{figure}

Based on this concept, Tangles build their consensus development based on the number of accepted predecessors in the Tangle, while not relying on a proof-of-work concept (e.g., in Blockchain) or on a certain time interval (e.g., as in CRDTs).
Figure~\ref{fig:blockchain:tangle} illustrates the differences between the Blockchain and Tangle technology.
If status changes happen, nodes can still operate on old data as they are not immediately evicted from the Tangle structure, allowing for situations in which the result of the operation might be validated.

\section{Tangle Centric Networking Architecture}
\label{sec:tangles}

The tight coupling of content identifiers challenges the introduction, and thus, the discovery of newly generated content in ICNs and compute-centric approaches as well (cf. Section~\ref{sec:intro}).
In order to address this problem, we propose Tangle Centric Networking (TCN).

TCN enhances the design principles of Interest-based ICN architectures (e.g., NDN) with the Tangle technology~\cite{popov2018tangle} to improve data retrieval of content in ICNs (see Section~\ref{sec:background}).
Like in ICNs, TCN uses naming schemes to request for NDOs and supports Manifests to query the network for large objects.
Large objects are divided into chunks, while each chunk  is associated with a content identifier, managed in a Manifest (e.g., FLIC~\cite{tschudin2016file}).


In TCN, the Manifest of each NDO is represented as a Tangle, associated with a content identifier that can be requested within an ICN.
Each Tangle represents a set of chunks of a NDO (or the entire NDO) in a structured way.
In case a change happens to a NDO, and therefore, to a set of chunks of the NDO, the characteristics of a Tangle allows network participants to acknowledge to the latest status of it and thus, to achieve consensus.
As a result, Tangles in TCN represent a common structure handling all versions of a particular NDO. 
It allows a consumer to selectively request for different versions of a NDO, without resolving several identifiers for each Manifest as required in FLIC.
Furthermore, the special update and synchronization characteristics of Tangles simplify the handling of changes of NDOs and is promising to improve the performance of updating Manifest.

\subsection{The Design of Tangles in TCN}
\label{sec:tangle:design}
A Tangle in TCN is represented by two elements: A \textbf{core element} -- representing a part of a NDO Manifest, containing a list of names of associated chunks, and an \textbf{acknowledgement element} -- representing a meta-information block to assist consensus. 

\begin{figure}
    \centering
    \includegraphics[width=.45\textwidth]{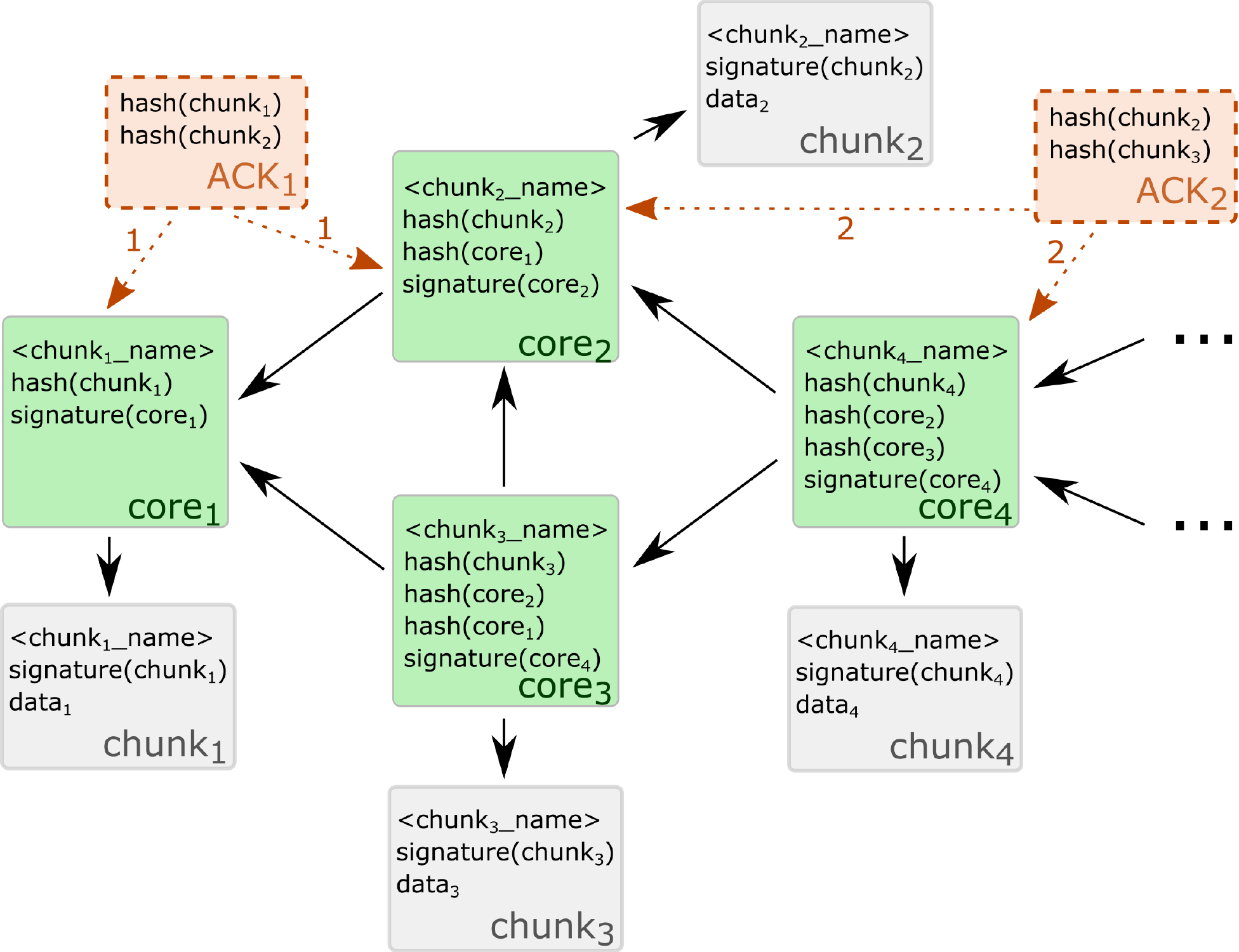}
    \caption{A Tangle consisting of \emph{core} (\textcolor{green}{green}) and \emph{acknowledgement elements} (\textcolor{red}{red}) which identify chunks (\textcolor{gray}{gray}) by hashes. Verification degree is the weight of the edges.}
    \label{fig:CoreElementTangle}
\end{figure}

The \emph{core element} stores the name as well as the hash of each associated chuck, used by nodes to query the network for data.
Each Tangle is signed by the producing node using digital signatures to ensure data integrity and authenticating the originator of the Tangle. 
In order to point to the next subsequent core element of the Tangle which contains the information to the next chunks, the element maintains the hash value of the previous core element and the hash value of at least one further chunk to form the initial Tangle. 
As a result, the Tangle created is represented as a DAG of elements (see Figure~\ref{fig:CoreElementTangle}).
The verification degree of a core element is the number of incoming edges of acknowledgements, while a Tangle \emph{tip} describes an element with no incoming edges.
By referring to the \emph{precedent} chunks instead of following ones (e.g., as designed in BlockChain~\cite{swan2015blockchain}), a data flow driven processing of Tangles is ensured by design.
Chunks within a Tangle are already available in the network to be requested by consuming nodes, while subsequent chunks are not yet created.


As core elements of a Tangle might get created and appended to the overall Tangle structure in irregular time intervals, a mechanism to model consensus is required. 
To model it, an acknowledgement element describes a meta-information block associated with each core element (see Figure~\ref{fig:CoreElementTangle}). 
Every time a core element is synchronized and accepted by another node in the network, an acknowledgement element for the particular core element is appended by the synchronizing node.
This way, a single node can only append one acknowledgement per appended block. Thus, with the knowledge how many nodes are synchronizing, a node knows when a consensus was established for a block. 
In TCN, we do not need a tip selection algorithm as in IOTA, since the places where a node can append data are already tightly restricted, since a new version can only be appended on the previous version
and acknowledgements can only be appended on the new block in addition to at least one randomly chosen other block. Therefore, the Tangle necessarily represents the logical structure of the data.\\

An example of a data exchange in TCN looks as follows: a client requests for a large NDO using an ICN Interest packet and the content identifier of the NDO. 
As it is a large NDO separated into several chunks, the client receives a Manifest file representing a part of a Tangle. 
The client uses the chunk information provided in the Manifest to request for each chunk of the NDO.
Tangles are stored in a dedicate Manifest Synchronization Table (MST) data structure instead of being handled as a typical content object in the ICN forwarding plane.
This has two reasons: (i) it allows to identify and operate on Tangles (e.g., synchronize its content due to updates), and (ii) avoids eviction of Tangles in the node cache according to its cache management policy.
When a forwarder stores a Manifest representing a core element of a Tangle in its \emph{MST}, it will append an acknowledgement block to each element that it could verify.
Verification takes place during forwarding the data chunk by comparing the hash value of the signature of the chunk packet with the signature hash value present in the Manifest.
The verification degree of a core element is the number of incoming edges, which represents the number of nodes which acknowledged to a core element.
The moment a forwarder will append an acknowledgement to a core element, the other nodes storing the manifest will synchronize the change. 
We refer to this situation as Tangle synchronization.
The more forwarder acknowledge to a core element the better in order to reach consensus on the Tangle.


\subsubsection{Linear vs. Tree Tangles}
Tangles can be modeled in different ways. 
Besides modeling a Tangle in a linear fashion where one element follows each other, the concept of Tangles also allows to model a hierarchical tree.

Instead of creating one ICN Manifest file describing the entire Tangle, e.g., by listing all core and acknowledgement elements in one Manifest, a publisher can create a Manifest file for each core and its associated acknowledgement element. 
As a result, elements of a Tangle are treated and transferred in the network as NDOs in ICN, allowing for new opportunities to handle data in TCN.

This is an important aspect, as additional nodes can contribute to the overall Tangle besides the original content publisher. 
For example in the \emph{electronic horizon} use case (cf. Section~\ref{sec:use_cases:automotive}), other nodes can improve the overall quality of the use case application by updating the Tangle with additional data sources dynamically (e.g., feeds of cameras deployed at buildings recording a hazardous situation in the vicinity). 
Being able to verify how many different nodes participated in the Tangle so far, either by creating new core elements or acknowledgements, a node knows how well certain NDOs are verified and which elements have been created by the originator and which parts have been appended by other nodes (e.g., Figure~\ref{fig:CoreElementTangle}).

\subsubsection{Handling Versions of Data in TCN}
According to the design rationale of immutable content identifiers in ICNs (e.g., Zhang et al~\cite{zhang2014named}), any change of the payload of a NDO results in a new version of the NDO, announced in the network using a new content identifier.
In order to be able to request for different versions, TCN's Tangles support versioning by appending new versions of individual chunks to the existing Tangle.
Instead of re-creating the entire Manifest, i.e., as in FLIC~\cite{tschudin2016file}, only the changed elements within a Tangle need to be updated, pointing to the new version of the NDO.
As a result, TCN allows to represent data in the network in a \emph{distributed database} fashion instead of a simple \emph{key-value store}.

However, a new version cannot be accepted without the consensus of other nodes in the network.
Whenever a change happens in a Tangle, a new version is created and appended to the Tangle, visible as history for all other nodes.
Similar like in Blockchain, all transactions on the Tangle are stored within the Tangle itself.
While a Blockchain represents a single decentralized data structure containing all content, replicated by every participant, TCN introduces decentralized data structures on each NDO.
As a result, the overhead of synchronizing new content is expected to be fewer compared to a Blockchain structure. 
Furthermore, Tangles provide more flexibility in selecting the content to be synchronized.
Staled information of a Tangle can be dropped by the publisher (e.g., non-valid version), instead of replicating all existing content as prescribed by a Blockchain.

\begin{figure}
    \centering
    \includegraphics[width=.45\textwidth]{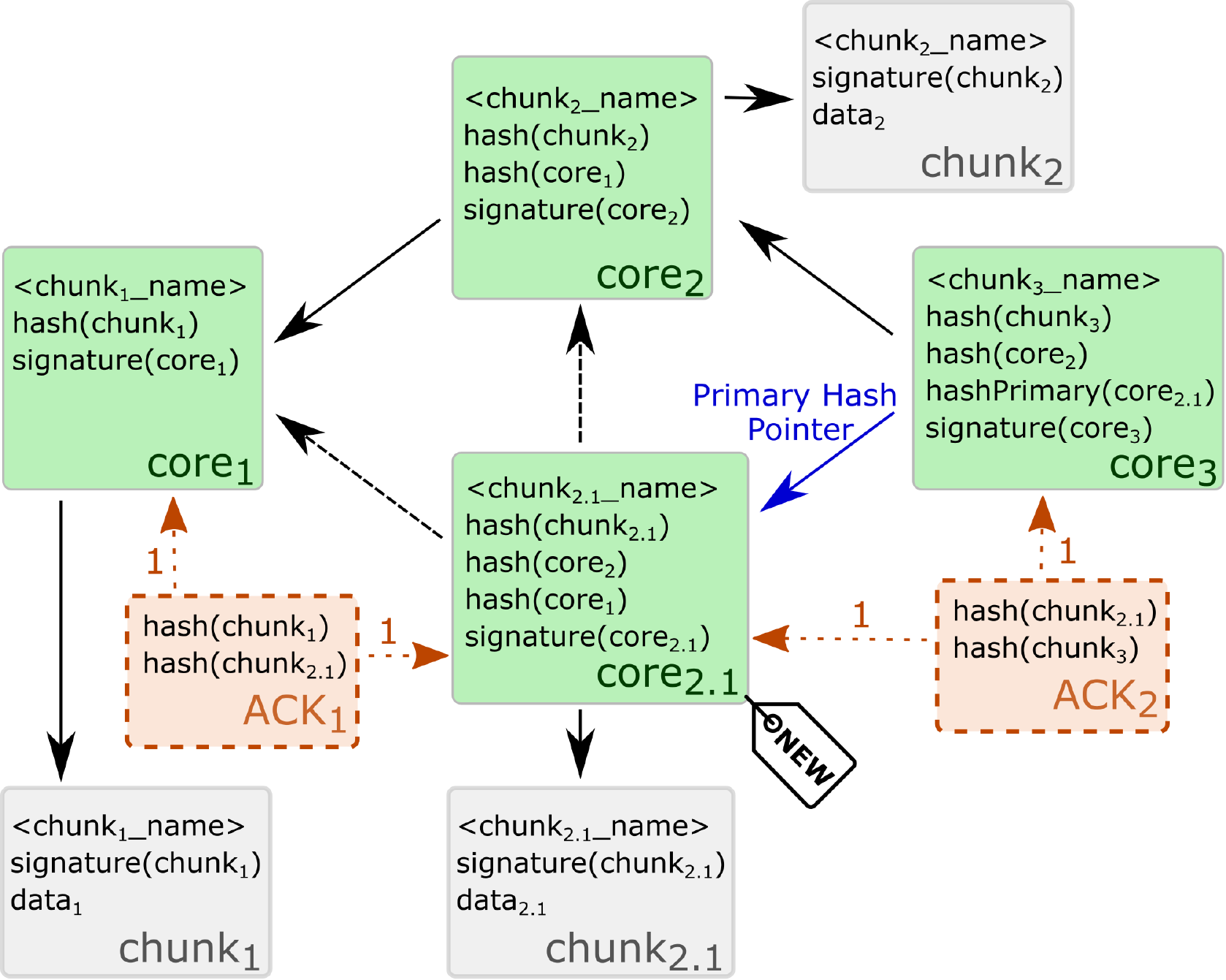}
    \caption{A new version $core_{2.1}$ is appended to the Tangle. Core elements have a primary hash pointer on the direct predecessor (here \textcolor{blue}{blue}).}
    \label{fig:CoreElementTangleAppend}
\end{figure}

In order to get a new version accepted in the network, a producer needs to select two core elements of a Tangle to append the new version to (cf. Figure~\ref{fig:CoreElementTangleAppend}).
We recommend that the primary element is selected in a semantic manner: if a new version of a chunk is added, it has to be appended to the block containing the previous version. 
If a chunk is replaced entirely by a new block, it has to be appended to the end of the Tangle representing the latest version.
The secondary element is selected randomly as each element at least has to point to two predecessors. 
A random selection minimizes the risk of circling pointers which increase the processing time of the entire Tangle.


\subsection{Tangle Representation in ICN}
\label{sec:Binary_Tangle_Representation}

In order to be able to transfer Tangles in the network, an explanation of representing Tangles in ICNs follows.
Instead of representing an entire Tangle as an application data unit in an ICN, each element of a Tangle is represented as a NDO.
It allows to fetch/synchronize parts of a Tangle instead of requesting for the entire one.

As mentioned in Section~\ref{sec:tangle:design}, a Tangle is given as a list of individual blocks, while each block contains at least two pointers to previous blocks in the structure -- the primary and secondary pointers.
The following listing illustrates the structure of a Tangle element in TCN:

{\scriptsize
\begin{lstlisting}
{
    type: <core, acknowledgement>
    chunkname: <name>,
    primaryhash: <hash>,
    secondaryhash: <hash>,
    ...
    signature: <signature_of_origin>
}
\end{lstlisting}}

When creating an acknowledgement element, there is no primary hash to be defined as the element represent meta-information.

Each element of a Tangle is transferred in the network as individual NDO and accessible in the underlying ICN using a content identifier.
In TCN, we propose the following naming structure to access an element of a Tangle:

{\footnotesize
\begin{verbatim}
    <NDO-Tangle-name>/<hash-value>/<version-number>
\end{verbatim}}

The structure allows to provide fine-grained access to a Tangle element, e.g., to a particular version.
The name of the actual chunk can be chosen according to the permitted name space.

Each time, a new element is added to the Tangle, a new version is created and appended as the \emph{tip}.
As \emph{tips} always represent the starting point for synchronizing a Tangle, the proposed naming convention to access the \emph{tip} of a Tangle is:

{\footnotesize
\begin{verbatim}
    <NDO-name>/<version-number>.
\end{verbatim}}

In order to ensure to receive the latest version from the origin, a consumer node has to specify a cache bypass signal as part of the request (e.g., ''\texttt{must-be-fresh}'' flag in NDN). 


\section{TCN - Synchronization of Tangles}
\label{sec:tangle_sync}

The synchronization of Tangles and their contents across multiple consuming nodes in the network describes a major aspect in TCN.
For example, if a new version of a NDO is created, both the Tangle and its elements as well as the consuming nodes have to be updated to the latest version.
However, as a Tangle represents a data structure distributed across the network, it describes a non-trivial task to update records as well as to update forwarding rules to reach the latest version.

\subsection{Bootstrapping and Synchronization}
\label{sec:tangle_sync:boot}

The synchronization is separated into two parts: (i) \emph{bootstrapping} -- the initial Tangle information are requested, and (ii) \emph{synchronization} -- partial or full synchronization of a Tangle with other nodes.

\subsubsection{Bootstrapping} The Tangle \emph{tips} are ideal to bootstrap a Tangle as they represent the elements without incoming edges.
By querying the network for the desired NDO using its content identifier, a consuming node will receive the \emph{tip} of the Tangle (cf. Section~\ref{sec:Binary_Tangle_Representation}).
Starting from the \emph{tips}, the node will traverse the Tangle to the first block, requesting for all predecessor elements as well as all chunks of the desired NDO using the (primary) hash pointers provided in each element.
While traversing backwards through the Tangle, a node can already fetch the latest versions of chunks.
By evaluating the acknowledgements of the Tangle, the node can decide to fetch further elements of the Tangle, or to drop these parts.
The node will proceed the initial bootstrapping as long as all elements are received.

\subsubsection{Synchronization} 
In order to get updates on the Tangle, the node will setup a synchronization of the \emph{tips}.
This can happen either by frequently poll for new versions, or using an ICN synchronization protocol such as PSync~\cite{Zhang:PSync:2017} or ChronoSync~\cite{Zhu:ChronoSync:2013}. 
Thereby, TCN benefits from the ICN in-network caching capability for which the network will automatically store replicas frequently in forwarders to increase the availability of data in case of node failures.
In case a new element is appended to a synchronized Tangle, either a new entry is added to the list of \emph{tips} or it will be linked to a synchronized \emph{tip}. 
As a result, any updates can be quickly identified and the Tangle is synchronized accordingly. 
It has to be mentioned that the synchronization interval to be selected for a Tangle is dependent on the application type. 
The frequent data is generated or requested by the application, the shorter synchronization intervals are required.

\subsubsection{Logical Tangle Representation for Special Interactions}
Special types of application interaction patterns, e.g., streaming of data such as video, are challenging to be realized as Tangles as the pure structure only allows traversing backwards.
To overcome this limitation, a node has to create a logical representation of the Tangle in the reverse order the Tangle is traversed from the tips to the start.
When a node starts synchronizing a Tangle, it will store the elements in the MST in reversed order as traversed from the tips to the start.
The logical representation allows to identify and request for the 'next' elements of the NDO to be streamed.

\subsection{Synchronization Chain - Update Forwarding Rules}

The synchronization of a Tangle across the network requires full interconnection of subscribing nodes as potentially each participating node can add new blocks to the Tangle.
This situation challenges the synchronization procedure in TCN as several updates might be received in a converging node.
To overcome this limitation, TCN establishes a logical synchronization chain between participating nodes.

After a consumer bootstrapped a Tangle, it will send out a special notification message into the network to discover at least one other node synchronizing on the same Tangle, e.g., using the postfix \texttt{findsyncpartner} in the content identifier.

{\footnotesize
\begin{verbatim}
    <Tangle-name>/<nonce>/findsyncpartner
\end{verbatim}}

If the notification message is overheard by a potential synchronization partner, the partner will respond back to the notification.
This response message is processed differently by the TCN forwarding pipeline by updating entries in the forwarding table to establish a bidirectional connectivity between these nodes.
As a result, request and response messages for the particular Tangle on the network interfaces are forwarded towards the synchronization partners.

\begin{figure}
    \centering
    \includegraphics[width=0.4\textwidth]{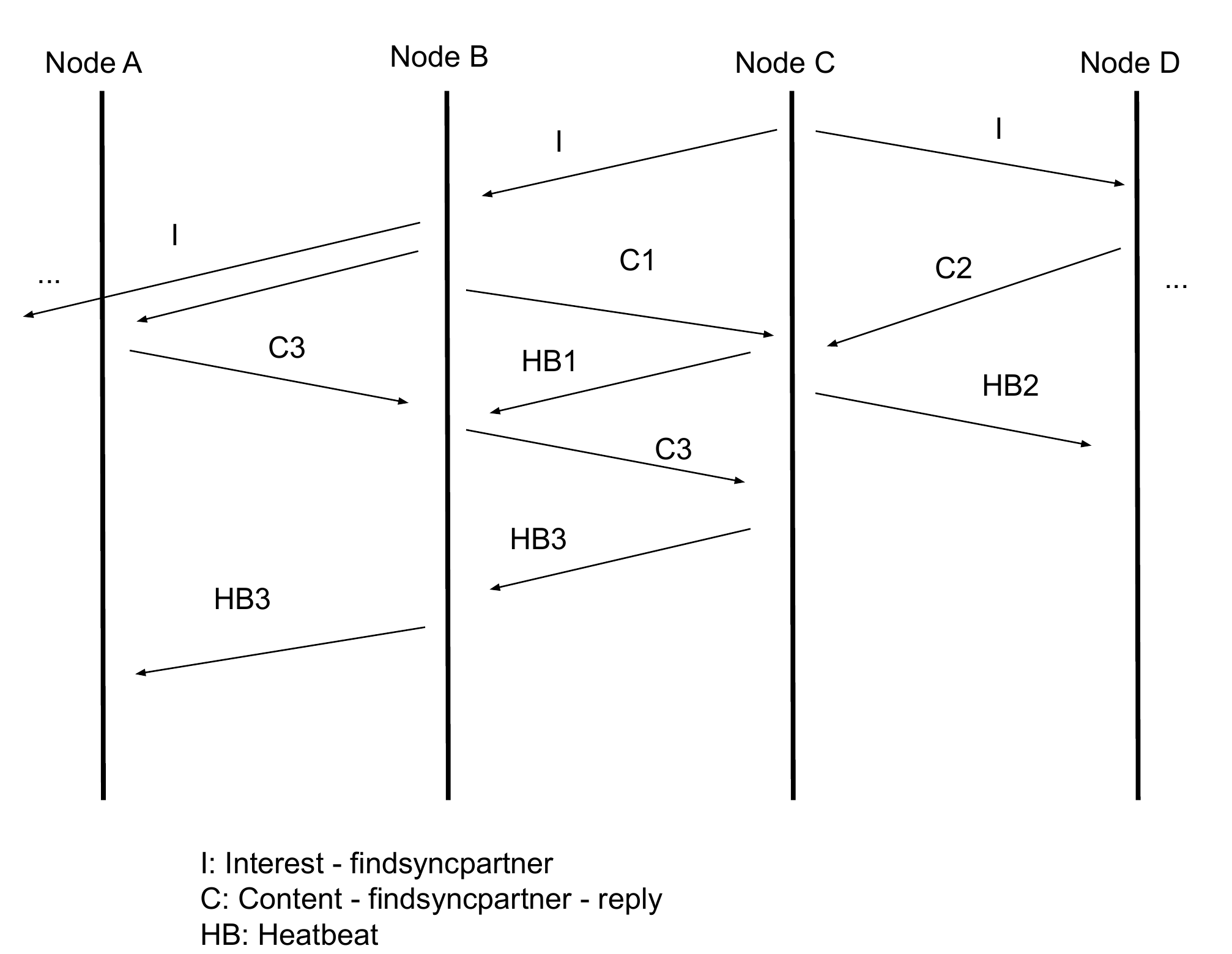}
    \caption{Scheme of the synchronization chain handshake procedure. First sending an interest to all matching FIB entries, next receiving a NDO to set time-limited forwarding rules, last completing the synchronization handshake and starting the heartbeat message procedure.}
    \label{fig:HandshakeProcess}
\end{figure}

To establish a bidirectional connectivity, we propose a three way handshake (cf. Figure~\ref{fig:HandshakeProcess}), so the new created FIB entries will only be persistent, if the initializing node confirms to the new routes.  
All forwarding entries created by TCN are time-limited to avoid security threats such as Denial-of-Service attacks. 
A heartbeat mechanism ensures that, in case of a timeout, FIB entries are removed accordingly.

After the synchronization routes are established successfully, the nodes can start to synchronize on the Tangle tips. In case a Tangle contains multiple tips received by a node, it has to merge the tips by eliminating duplicated entries in the list of tips.

In case a connection loss happened and the disconnected node extends the Tangle with new elements (e.g., the vehicle drives into a tunnel having no connection to the communication infrastructure), it will restart the bootstrapping of the Tangle after re-connecting to the network. Other synchronization partners will receive the changes and decide whether to accept or reject to the latest changes using acknowledgements.




\section{Evaluation} 
\label{sec:evaluation}

In order to evaluate the scalability of TCN, the networking simulator NS-3~\cite{ns3:2018} and the NDN specific protocol stack bundle ndnSIM 2.7~\cite{ndnSim:2017} are used to create the simulation.
Dependent on the number of synchronization nodes, the time required to synchronize to parts as well as a full Tangle is compared against the FLIC Manifest proposal.

\subsection{Synchronization}
\begin{figure}
    \centering
    \includegraphics[width=.4\textwidth]{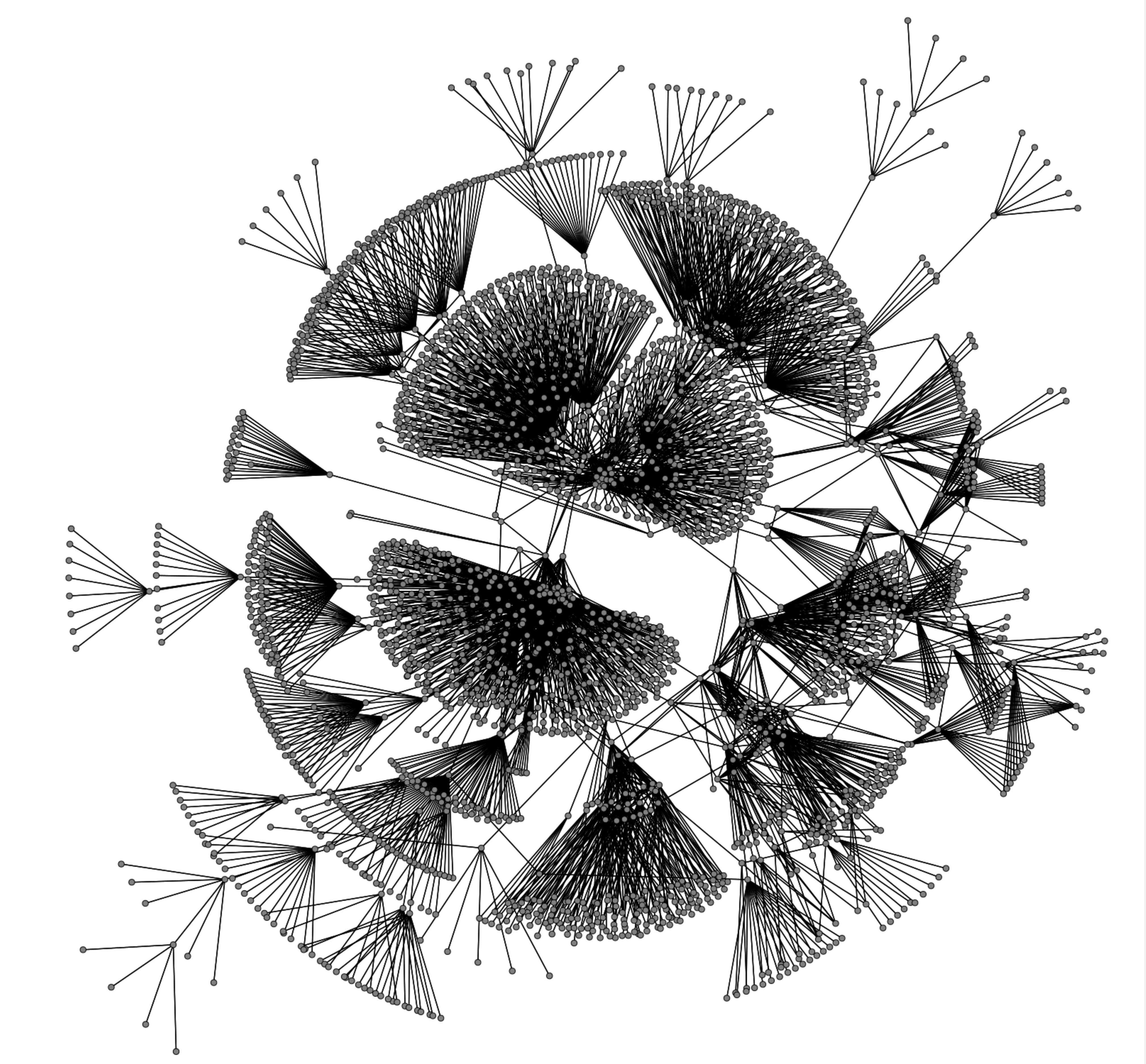}
    \caption{Rocketfuel Topology with 2766 nodes, visualized with pyviz in ndnSIM.}
    \label{fig:rocketfuel}
\end{figure}

As topology for our simulation we use the rocket-fuel typology~\cite{spring2002measuring}\footnote{more specific the \texttt{1221.cch} based on \url{https://github.com/cawka/ndnSIM-sample-topologies}} including 2766 nodes with a bandwidth capacity between $2$MBps and $10$MBps, representing a hierarchical network as shown in Figure~\ref{fig:rocketfuel}. We run the evaluation $100$ times and every time we initialized the bandwidth capacity of the links differently.

To evaluate the scalability of TCN, we choose randomly $n$ nodes (range: $n=4$ up to $n=256$) to be involved in the procedure. 
The size of the NDO to be modeled as a Tangle is $1$GB, while each chunk is of size $50$KB. 
Based on these parameters, the Tangle to be simulated consists max. of $20973$ blocks. 
In the Tangle bootstrapping, all $n$ nodes synchronize the content of the Tangle. 
Afterwards, nodes will start to append new chunks of max. size $200$MB to the Tangle (max. $4096$ additional blocks in total).
We measure the time taken to synchronize the entire Tangle at every participating node. 
Moreover, we request the actual data and measure the time until all nodes have fully synchronized to the desired data.
The bandwidth is randomly chosen per link, between $2$MBps and $10$MBps.

\begin{figure}
    \centering
    \includegraphics[width=.4\textwidth]{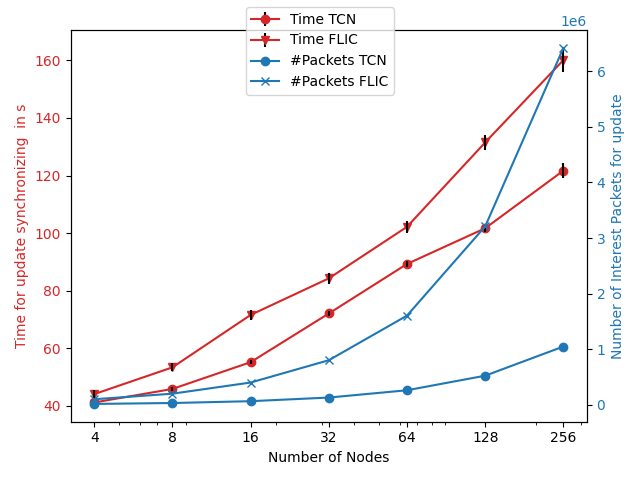}
    \caption{Result of full synchronization using TCN and FLIC (100 runs with random bandwidth initialization).}
    \label{fig:sync}
    \vspace{-0.4cm}
\end{figure}

We simulate the scenario either using FLIC and a full-synchronization of a Tangle. 
The results are illustrated in Figure~\ref{fig:sync}. It can be seen that FLIC requests the entire Manifest again when a change happend in the NDO. 
When looking at the synchronization time, it can be seen that TCN has a rather constant time saving compared to full synchronization in FLIC.

This is due to the fact that a fewer number of Manifest packets have to be transferred in the network as we append elements to the Tangle, while FLIC requires a re-creation of the entrie Manifest structure.
The experiment shows, that an efficient Manifest structure can help to speed up real world applications.
The synchronization time has a roughly linear grow despite the number of nodes is exponential growing. 
This can be explained by the intrinsic multicast and in-network caching capabilities of NDN as forwarding nodes in the simulated topology will respond to synchronization requests with cached data.

\begin{figure}
    \centering
    \includegraphics[width=.4\textwidth]{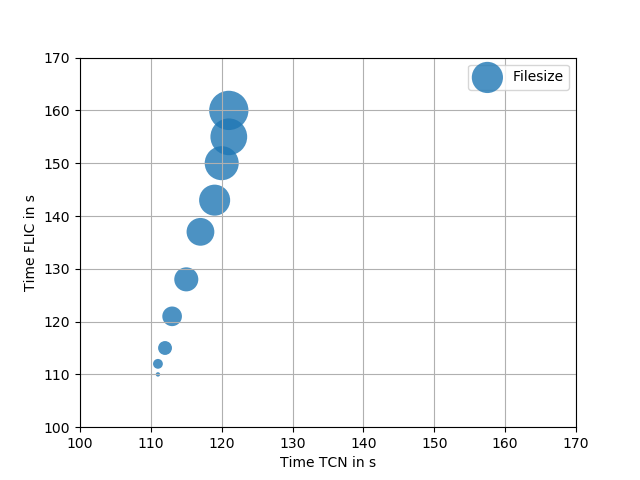}
    \caption{Result of full synchronization using TCN and FLIC with different file sizes.}
    \label{fig:Scale}
    \vspace{-0.5cm}
\end{figure}

In a second experiment we use the same setting, but we want to analyze the scaling behavior of TCN vs FLIC. Therefore, we change the size of the initial data to $10$, $50$, $100$, $200$, $300$, $400$, $500$, $600$, $700$, $800$, $900$ and $1000$MB
and we observe the time how long TCN and FLIC need to synchronize additional $200$MB.
The result is shown in Figure \ref{fig:Scale}.

We see that for small data ($10$, $20$, $50$MB) TCN and FLIC have a rather similar performance. However, with increasing file sizes, full synchronization with FLIC as manifest becomes slower than TCN due to the overhead of requesting the entire manifest compared to TCN only fetching the new blocks. Thus, we find that with increasing file size TCN scales very well. This is important, since when the amount of synchronized data grows, the synchronization overhead should not increase for optimal performance.



\section{Discussion}
\label{sec:discussion}

In IOTA, Tangles offer a lightweight solution to handle data of IoT scenarios in a distributed manner, while not relying on concepts such as proof-of-work or a transaction fee as Blockchain.
By providing additional features such as consensus between partner nodes as well its conceptual fit to the proposed structure of Manifests as in FLIC~\cite{tschudin2016file}, TCN overcomes the limitations of tight name-to-data coupling.
These properties makes TCN an interesting candidate for distributed applications deployed especially in edge computing environments.
However, there are additional aspects of TCN not addressed so far.


\subsection{TCN -- Security \& Trust}
The current design of TCN allows easy participation of nodes on the Tangle structure.
Every node which is able to synchronize a Tangle is able to append data to it, even malicious one, due to a missing concept for authorization. 
However, similar as in Blockchains, Tangles establish a trust model based on consensus. 
Data will only be synchronized if an element is acknowledge by subsequent elements.  
Elements which get significantly less acknowledgements will be dropped from the Tangle after some time (Tangle convergence), while verified elements remain in the Tangle. 

As in ICN every node is required to sign a content packet, signatures within the packets might be used to authenticate a node against a Tangle.
Additional mechanisms to handle permissions to operate on the Tangle can help to prevent a node publishing malicious data.


\subsection{TCN -- Intelligent Cache Management}

The information modeled within a Tangle can be used for cache decision making.
A node synchronizing a Tangle can anticipate the next data chunk to be requested by a consumer node, pre-fetching and storing it within its local cache pro-actively.
As a result, TCN forwarding nodes turn into data repositories.
Such feature might be useful in time-sensitive use case scenarios such as connected vehicles~\cite{scherb2019data}.
For example, proactively storing data at edge nodes, e.g., traffic information about a hazardous situation ahead, improves data dissemination in mobile scenarios.
Furthermore, as Tangles allow to append data to an existing structure, vehicles can append own generated data or even carry data to areas not covered by a communication infrastructure, and thus, contribute to  improve the quality of the distributed application. 


\subsection{TCN -- Flexible Name-to-Data Binding}

ICNs tightly couple content identifier and actual data.
The mandatory signing procedure of ICNs ensures that these bindings can be verified at client side (or even forwarder) by relying on a Public Key Infrastructure (PKI).
New or updated content has to be published using a new identifier.
Especially, Manifest solutions like FLIC require to re-create its entire structure in case of any changes.
TCN revises the tight coupling as Tangles allow to extend the existing structure by adding new content to the Tangle on demand.
The acknowledgement mechanism of Tangles helps to establish consensus on valid elements, while less valid content will be evicted from the Tangle from time to time.
As a result, data can be enhanced not only by the originator of the data which is essentially different compared to plain NDN solutions.
To this end, TCN moves management complexity from the application level into the Tangles enabling the support of cooperative application.

\subsection{TCN -- Towards Compute-Centric Architectures}

The introduction of TCN have shown improvements via network level consensus and efficient management of data chunks (cf. Section~\ref{sec:evaluation}).
However, the concepts of Tangles are promising for compute-centric architectures such as Named Function Networking (NFN)~\cite{sifalakis2014information}, 
named function as a service (NFaaS)~\cite{krol2017nfaas}, or Compute First Networking (CFN)~\cite{krol2019compute} as well.
For example, the \emph{electronic horizon} application logic can be modeled as a compute graph using Tangles.
By doing so, the Tangle can be synchronized between execution nodes to coordinate the execution of computations across cloud and edge environments, and thus, ensure to fulfill application demands. 
Further application domains which might benefit from such concept include distributed database systems, or distributed chat applications like Scuttlebutt~\cite{tarr2019secure, scherb2021scoiot}.
We briefly discuss the potentials of the future work on Tangle Function Networking (TFN) -- an extension of TCN supporting the coordination of computation state across multiple execution nodes in a distributed and transparent fashion.

\subsubsection{Tangle Function Networking}
NDN has been proven by several applications to be a good basis for network computing. NFN~\cite{sifalakis2014information, scherb2016packet}, 
NFaaS~\cite{krol2017nfaas}, CFN~\cite{krol2019compute}, etc are examples for 
network computing frameworks for NDN.
When looking into the concepts of NFN and NFaaS, it can be seen that both support the execution of stateless functions as in the ''function-as-a-service'' paradigm.
Besides stateless functions, CFN supports the execution of stateful Actors by synchronizing compute state using CRDTs.
Instead of using CRDTs, Tangles can be used for the following reasons: (i) allow for collaborative resolution of execution nodes, and (ii) allow to dynamically extend the graph by other nodes during runtime (e.g., to store and synchronize local variables and intermediate results).

\begin{figure}
    \centering
    \includegraphics[width=0.45\textwidth]{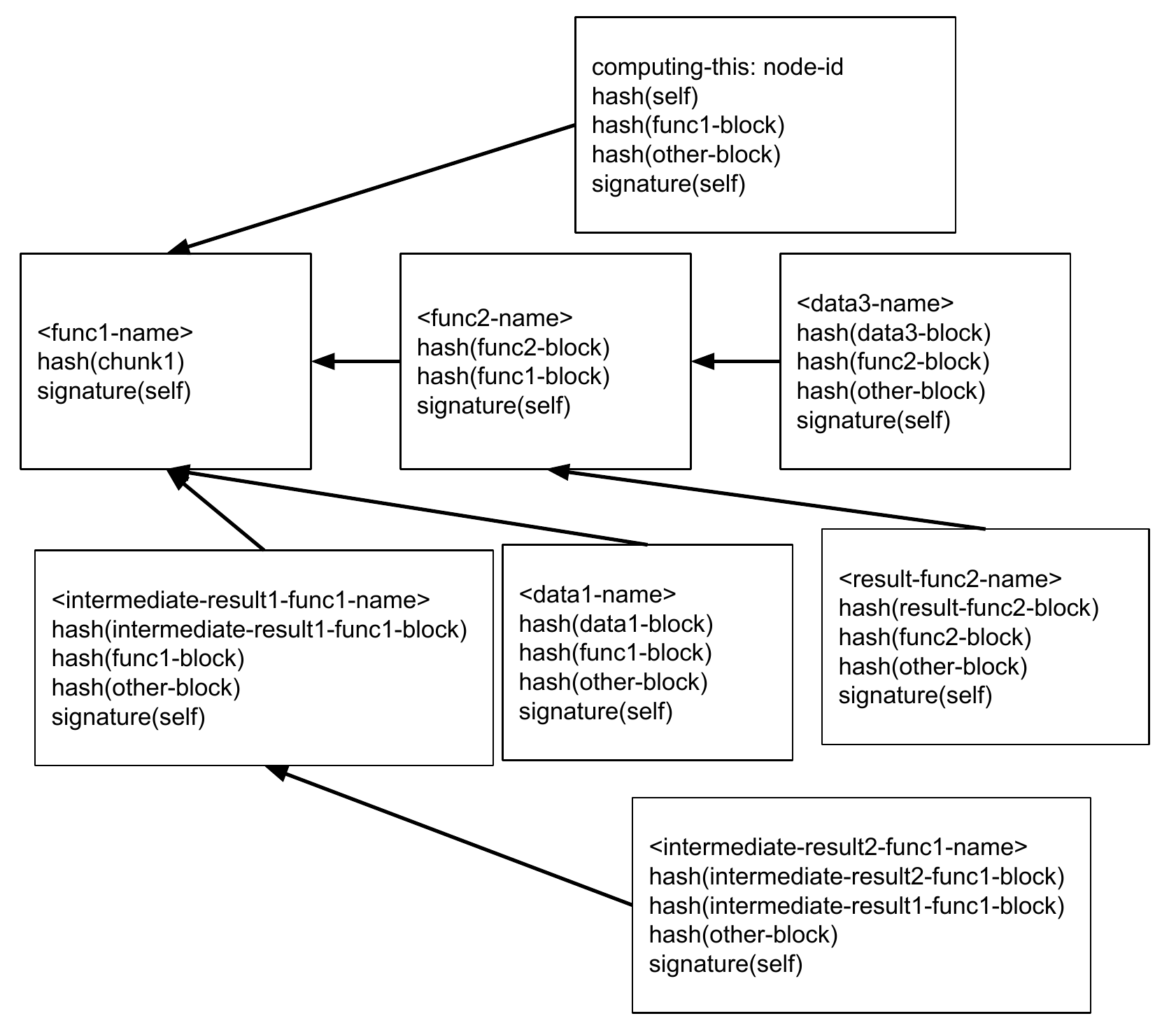}
    \caption{Compute Graph of the function call \texttt{func1(func2(data3),data1)} in TFN. This figure shows only logical compute graph excluding the second hash reference of the tangle.}
    \label{fig:TFNComputeGraph}
    \vspace{-0.5cm}
\end{figure}

A Tangle in TFN consists of two blocks: \emph{functions}, and \emph{data} (input, compute state or compute result), therefore, supports both stateless functions and stateful computations (cf. Figure~\ref{fig:TFNComputeGraph}). 
Each of these blocks are represented as individual elements in the Tangle.
When an execution node parses the Tangle, it can decide to execute a function of the graph.
Pointers to additional information such as the location of the function executable, as well as further data is available in the Tangle as well.
In case a compute result is available, the Tangle is updated accordingly that subsequent execution nodes can query quickly the network for the result.
Furthermore, modeling a compute graph using Tangles allows for parallel processing of the entire graph.
In case multiple execution nodes are involved in the overall computation, a node takes over responsibility to execute a function by acknowledging the execution in the Tangle, and synchronized partners will be notified accordingly. 
Furthermore, it limits the probability that a function within the Tangle is executed in parallel by different nodes.

As we move toward a distributed edge computing environment, the concept of TCN is promising for compute-centric networking architectures~\cite{scherb2018smart, scherb2019execution}.
It allows the compute network to change or extend computation graphs during the runtime, bringing more flexibility than all other computation frameworks introduced so far~\cite{scherb2017execution}.
For example in the \emph{electronic horizon} scenario (cf. Section~\ref{sec:use_cases}), to warn the driver about a hazardous situation on the junction ahead, nodes can append image processing computations of a camera deployed close to the junctions to the compute Tangle or append the processing result to the Tangle as additional input for the computation~\cite{scherb2018resolution, grewe2018network}.
Another promising feature, especially in mobile scenarios is the capability that Tangle-based computations can be continued offline in case of connection loss.

\section{Conclusion}
\label{sec:conclusion}

In this paper, we present the challenge of tight name-to-data binding in ICNs, especially for scenarios in which Manifest solutions are used.
To address the limitations of the tight coupling, we present Tangle Centric Networking, a network level consensus and synchronisation extensions for ICNs.
In TCN, the Manifest of each NDO is represented as a Tangle, that enables the network to add or update content across multiple consumers in a distributed fashion.

We implemented the concept of TCN in simulation and compared it against the ICN Manifest solution FLIC.
The results show that Tangles provides an efficient way to synchronize immutable and mutable state over the network, while the additional synchronization overhead is negotiable when popular large files frequently change in the network.
Therefore, TCN and its Tangle concept improves the performance of data retrieval for these types of data in an ICN.

As we move toward a distributed edge computing environment, we discuss the concept of TCN as a promising extension to manage compute-centric networking architectures more efficiently.
Future work of TCN has to address the presented open directions, especially these points regarding security and trust management.

\bibliographystyle{ACM-Reference-Format}
\bibliography{bib/ms}

\end{document}